%% file: proctop.tex
\documentclass[12pt]{article}
\usepackage{epsfig}

\input econfmacros.tex
\def\Title#1{\begin{center} {\Large {\bf #1} } \end{center}}

\begin{document}

\Title{Top-pair and $tW$ production at approximate N$^3$LO}

\bigskip\bigskip

\begin{raggedright}  

{\it Nikolaos Kidonakis\footnote{Talk presented at the APS Division of Particles and Fields Meeting (DPF 2017), July 31-August 4, 2017, Fermilab. C170731}\\
Department of Physics\\
Kennesaw State University\\
Kennesaw, GA 30144, USA}
\bigskip\bigskip
\end{raggedright}

\section{Introduction}

I present approximate N$^3$LO theoretical results for top-antitop pair production, and for single-top production in association with a $W$ boson. The higher-order corrections are from soft-gluon radiation, which is dominant near partonic threshold. I present results for total cross sections as well as transverse-momentum and rapidity distributions of the top quark, and compare with data at LHC energies.

\section{Top-pair production}

The top quark is the heaviest elementary particle to have been discovered
to date. Because of its large mass, its near-threshold production at current 
colliders receives large radiative corrections from soft-gluon emission.
These corrections have long been known to be significant for $t{\bar t}$  
production, and they approximate exact results, when known, very accurately. 

Resummation of the double-differential cross section at next-to-next-to-leading logarithm (NNLL) accuracy in moment space was derived in Ref. \cite{NNLL} using the two-loop soft anomalous dimension \cite{NNLL,2loop}. Fixed-order expansions of the resummed cross section in momentum space bypass the problem of using a prescription for divergences, and they provide excellent and reliable predictions for the higher-order corrections. General expressions for the expansions have been derived and used for various processes at NNLO \cite{aNNLO} and N$^3$LO \cite{aNNNLO}. Approximate N$^3$LO (aN$^3$LO) predictions for double-differential cross sections in $t{\bar t}$ production have appeared in Ref. \cite{topaN3LO}. These aN$^3$LO corrections are needed for precision physics as they considerably enhance the total cross section and differential distributions.

An interesting question in the study of $t{\bar t}$ production is the effect of scale choice on the top-quark $p_T$ distributions. Traditionally, two central choices have been made for the factorization and renormalization scales: $\mu=m$, the top quark mass; and $\mu=m_T=(p_T^2+m^2)^{1/2}$, the transverse mass. The difference in the $p_T$ distributions using the two scales thus grows with $p_T$.

\begin{figure}[htb]
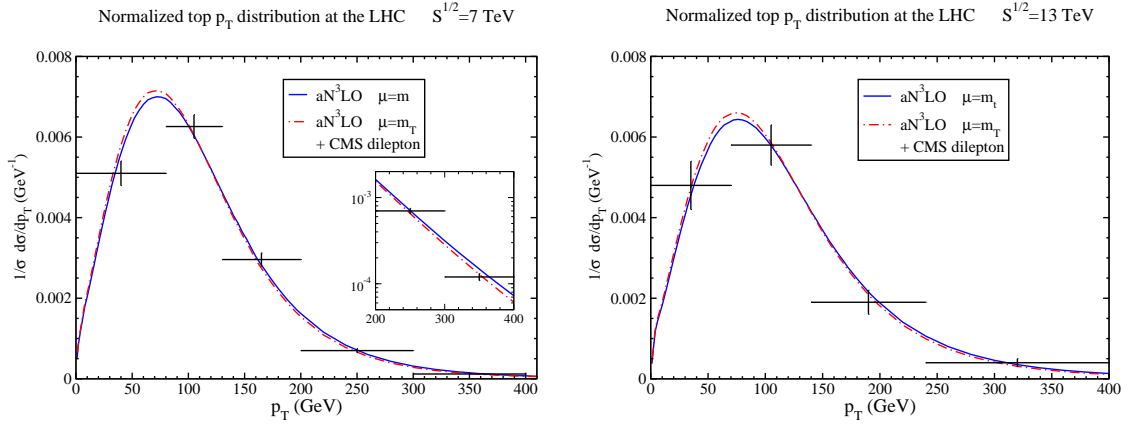

\begin{center}
\epsfig{file=pt7lhcnormCMSdileptplot.eps,height=2.2in}
\hspace{3mm}
\epsfig{file=pt13lhcnormCMSdileptplot.eps,height=2.2in}
\caption{Top quark aN$^3$LO normalized $p_T$ distributions compared with CMS dilepton data at (left) 7 TeV \cite{CMSpt7dilepton} and (right) 13 TeV \cite{CMSpt13dilepton} LHC energies.}
\label{pt7and13}
\end{center}
\end{figure}

In Fig. \ref{pt7and13} we display the top quark normalized $p_T$ distributions at 7 TeV (left plot) and 13 TeV (right plot) LHC energies. We compare with data from CMS in the dilepton channel at 7 TeV \cite{CMSpt7dilepton} and 13 TeV \cite{CMSpt13dilepton}. We find excellent agreement in both cases, especially with the choice $\mu=m_T$, which better describes the data at high $p_T$. We have used MMHT2014 pdf \cite{MMHT2014}; the results with CT14 pdf \cite{CT14} are similar.

\begin{figure}[htb]
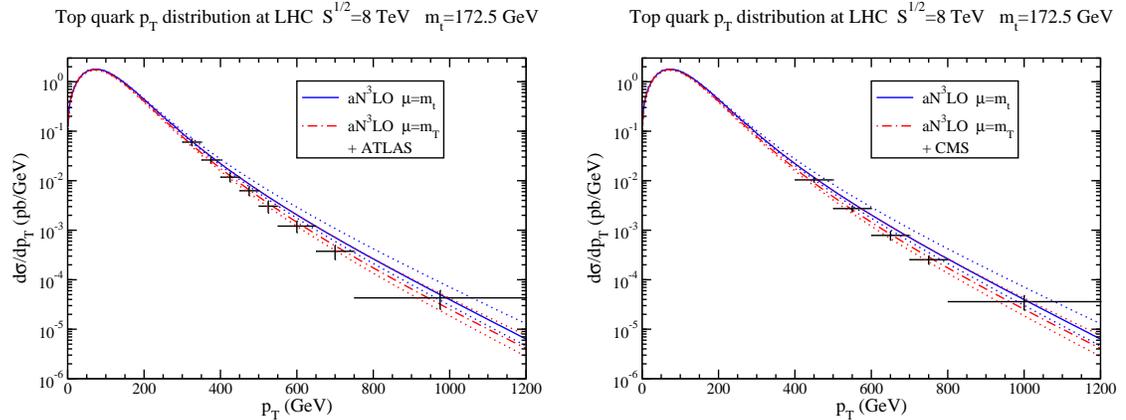

\begin{center}
\epsfig{file=pt8lhcaN3LOboostATLASplot.eps,height=2.2in}
\hspace{3mm}
\epsfig{file=pt8lhcaN3LOboostCMSplot.eps,height=2.2in}
\caption{Top quark aN$^3$LO $p_T$ distributions compared with (left) ATLAS \cite{ATLAS8boost} and (right) CMS \cite{CMS8boost} lepton+jets data at 8 TeV LHC energy.}
\label{pt8}
\end{center}
\end{figure}

In Fig. \ref{pt8} we display the boosted top quark $p_T$ distributions, with 
theoretical uncertainty, at 8 TeV LHC energy, and compare with lepton+jets data 
from ATLAS \cite{ATLAS8boost} (left plot) and CMS \cite{CMS8boost} 
(right plot). Again, the calculations with the choice $\mu=m_T$ 
better describe the data at high $p_T$.

\section{$tW$ production}

The associated production of a top quark and a $W$ boson, via the partonic process $bg \rightarrow tW^-$, was studied at NNLL accuracy in Ref. \cite{NKtW} using results for the two-loop soft anomalous dimension. Approximate NNLO (aNNLO) predictions for the total $tW^-$ production cross section were given in \cite{NKtW}. The cross section for ${\bar t} W^+$ production is the same. 

Top-quark $p_T$ distributions at aNNLO for this process were given in Ref. \cite{pTsingletop}. More recently, aN$^3$LO results for the total cross section and the top $p_T$ and rapidity distributions in $tW$ production were presented in Ref. \cite{tWaN3LO}.

\begin{figure}[htb]
\begin{center}
\epsfig{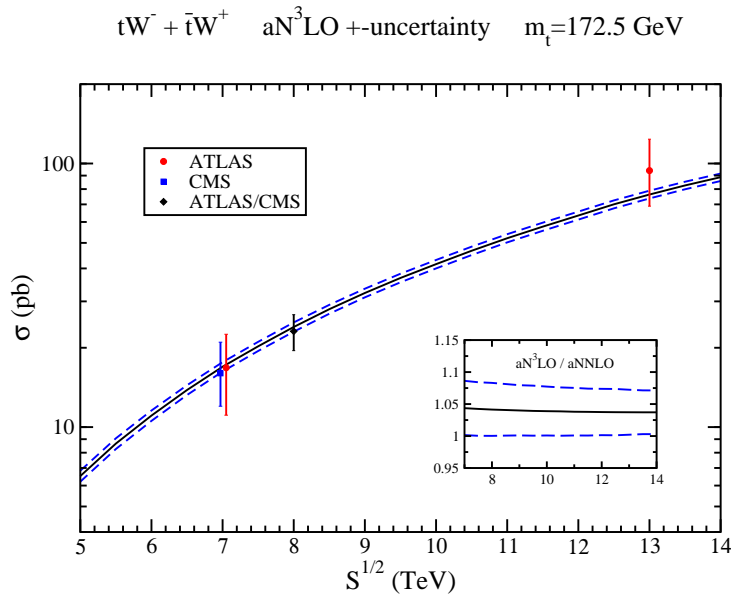}
\caption{Total aN$^3$LO cross section for $tW$ production at LHC energies compared with data \cite{ATLAStW7,CMStW7,ATLASCMStW8,ATLAStW13} from ATLAS and CMS.}
\label{tWS}
\end{center}
\end{figure}

In Fig. \ref{tWS} we show the total aN$^3$LO cross section, together with theoretical uncertainty, for $tW$ production and compare with LHC data. The theoretical predictions are in very good agreement with the data from ATLAS \cite{ATLAStW7} and CMS \cite{CMStW7} at 7 TeV, an ATLAS/CMS combination at 8 TeV \cite{ATLASCMStW8}, and ATLAS at 13 TeV \cite{ATLAStW13}. 

The inset plot in Fig. \ref{tWS} shows the aN$^3$LO/aNNLO ratio. It is clear that the third-order soft-gluon corrections are non-negligible.

\begin{figure}[htb]
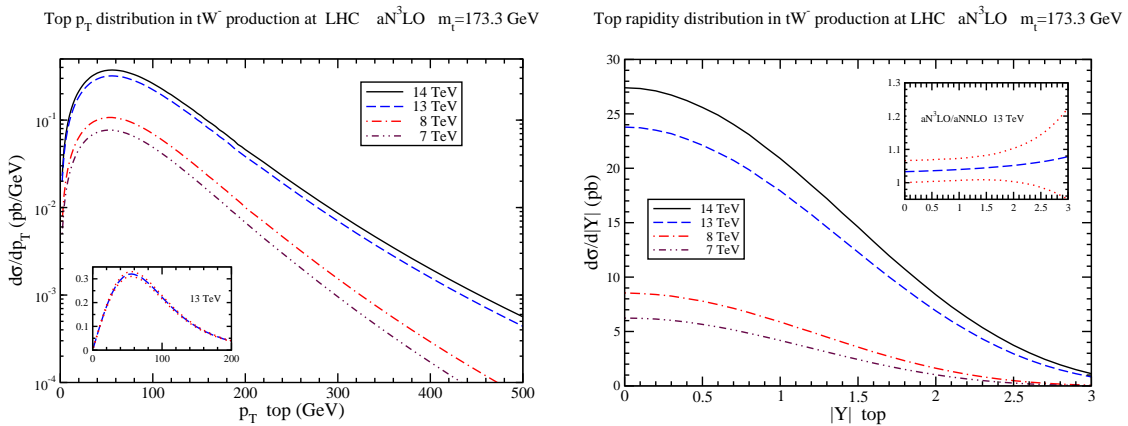

\begin{center}
\epsfig{file=pttoptWplot.eps,height=2.2in}
\hspace{3mm}
\epsfig{file=yabstoptWplot.eps,height=2.2in}
\caption{Top quark aN$^3$LO $p_T$ and rapidity distributions in $tW$ production.}
\label{ptytW}
\end{center}
\end{figure}

In the left plot of Fig. \ref{ptytW} we display the aN$^3$LO top quark $p_T$ distributions  in $tW$ production at LHC energies. The inset plot shows the distribution at 13 TeV energy together with the theoretical uncertainty.

In the right plot of Fig. \ref{ptytW} we display the aN$^3$LO top quark rapidity distributions in $tW$ production at LHC energies. The inset plot shows the 
aN$^3$LO/aNNLO ratio, with theoretical uncertainty, at 13 TeV. We observe that the soft-gluon corrections are substantial, particularly at large values of rapidity.

\section*{Acknowledgements}
This material is based upon work supported by the National Science Foundation 
under Grant No. PHY 1519606.

\end{document}

%% file: econfmacros.tex
\def\beq{\begin{equation}}
\def\eeq#1{\label{#1}\end{equation}}
\def\eeqn{\end{equation}}

\def\beqa{\begin{eqnarray}}
\def\eeqa#1{\label{#1}\end{eqnarray}}
\def\eeqan{\end{eqnarray}}

\let\bar=\overbar

\def\Dslash{\not{\hbox{\kern-4pt $D$}}}
\def\dslash{\not{\hbox{\kern-2pt $\del$}}}

\def\msb{{\bar{\ssstyle M \kern -1pt S}}}

